\documentclass[aps,pra,twocolumn,showkeys,longbibliography,superscriptaddress,showpacs]{revtex4-1}
\usepackage[utf8]{inputenc} 

\usepackage{amsmath,amsthm,paralist}
\usepackage{amsfonts}
\usepackage{amssymb}
\DeclareMathOperator*{\argmax}{arg\!\max}

\usepackage{times}

\usepackage{algcompatible}
\usepackage{algorithm}

\usepackage[english]{babel}
\usepackage{subcaption}
\usepackage{diagbox}

\usepackage{pgfplots}
\pgfplotsset{compat=1.14}

\usepackage{tikz}
\usetikzlibrary{arrows.meta,calc,chains,intersections,positioning,arrows}

\usepackage{rotating}

\newcommand{\nix}[1]{}
\newcommand{\ket}[1]{|#1\rangle}

\usepackage[hidelinks,
    colorlinks = true,
    urlcolor = violet, 
    linkcolor = cyan, 
    citecolor = red 
]{hyperref}

\usepackage{graphicx}
\usepackage{dcolumn}
\usepackage{bm}

\begin{document}

\preprint{AIP/123-QED}

\title{Neural Decoder for Topological Codes using Pseudo-Inverse  of Parity Check Matrix}

\author{Chaitanya Chinni}
\affiliation{FoodStreet.in, Chennai 600 042, India}%
\affiliation{YNOS Venture Engine CC Pvt. Ltd., Chennai 600 113, India}%
\author{Abhishek Kulkarni}%
\author{Dheeraj M. Pai}%
\author{Kaushik Mitra}%
\author{Pradeep Kiran Sarvepalli}%
\affiliation{ 
    Department of Electrical Engineering, Indian Institute of Technology Madras, Chennai 600 036, India
}%

\date{\today}

\begin{abstract}
Recent developments in the field of deep learning have motivated many researchers to apply these methods to problems in quantum information. 
Torlai and Melko first proposed a decoder for surface codes based on neural networks. 
Since then, many other researchers have applied neural networks to study a variety of problems in the context of decoding.
An important development in this regard was due to Varsamopoulos {\em et al.} who proposed a two-step decoder using neural networks. 
Subsequent work of Maskara {\em et al.} used the same concept for decoding for various noise models.
We propose a similar two-step neural decoder using inverse parity-check matrix for topological color codes. 
We show that it outperforms the state-of-the-art performance of non-neural decoders for independent Pauli errors noise model on a 2D hexagonal color code.
Our final decoder is independent of the noise model and achieves a threshold of $10 \%$.
Our result is comparable to the recent work on neural decoder for quantum error correction by Maskara {\em et al.}. 
It appears that our decoder has significant advantages with respect to training cost and complexity of the network for higher lengths when compared to that of Maskara {\em et al.}.
Our proposed method can also be extended to arbitrary dimension and other stabilizer codes.
\end{abstract}

\maketitle

\section{\label{sec:Introduction} Introduction}
In quantum computers basic unit of information is a qubit. Qubits are highly susceptible to noise. Hence to protect the information, we use quantum codes.
A very popular class of quantum codes for protecting information are topological quantum codes. In this paper we focus on a subclass of topological codes in two spatial dimensions called color codes \cite{bombin06}.
To correct the impact of noise on the encoded information we would need a decoder. Novel  decoding algorithms for 2D color codes have been proposed  earlier in \cite{wang2009graphical,sarvepalli12,delfosse14,bombin2012universal}.
However, these are not optimal and do not meet the theoretical bounds for performance.
Furthermore, designing decoders for non-Pauli noise is a challenging problem. 

Recent developments in the fields of machine learning (ML) and deep learning (DL) have motivated many researchers to apply these methods 
to decoding quantum codes. 
Torlai and Melko were the first to propose a decoder for surface codes based on neural networks \cite{PhysRevLett.119.030501}. 
Since then, many other researchers have applied neural networks to study a variety of problems in the context of decoding 
~\cite{PhysRevLett.119.030501,varsamopoulos2017decoding,Krastanov2017,baireuther2018neural,chamberland2018deep,davaasuren2018general,jia2018efficient,Breuckmann2018scalableneural,Baireuther2018machinelearning,maskara2018advantages}.

In this paper we only focus on decoding of color codes using neural networks. 
Early work based on neural networks attempted to the solve the problem using using neural networks entirely. 
These did not beat the non-neural methods. 
An important development in this context was due to \cite{varsamopoulos2017decoding} who proposed a combination of neural networks and non-neural decoders. 
More precisely, they have a two-step decoder where in the first-step, they estimate an pure-error and in the second-step, they use a neural network which estimates the logical.
In their recent work \cite{varsamopoulos2018designing}, they mention that any simple decoder can be used in the first-step.
The authors of \cite{davaasuren2018general} claim that the work of \cite{varsamopoulos2017decoding} is a special case of their generalized framework of building neural networks for decoding stabilizer codes.
The works of \cite{baireuther2018neural,chamberland2018deep} attempt to use neural networks for fault-tolerant setting.
The most relevant work to ours is \cite{maskara2018advantages} in which a similar combination of two decoders is employed to conclusively demonstrate the usefulness of neural decoders. 
They proposed a neural decoder with progressive training procedure that outperformed previously known decoders for 2D color codes. 

In this work, we propose a similar two-step neural decoder for color codes and study its performance for the hexagonal color code on the torus. 
We propose two variations, one which achieves a threshold of $10 \%$ and another with an important modification that achieves a near optimal threshold for independent bit-flip/phase-flip noise model. 
This modification can be incorporated in other neural network based decoders and could be of potentially larger importance.
The main challenge involved with neural networks is determining the correct architecture in order to improve the overall threshold. 
We model our non-neural decoder in a simple way and show the advantages of doing so with the improvement in performance of the neural decoder, the reduction in cost of training and scaling associated with the distance of the code.
Our main contributions are,
\begin{compactenum}[1)]
    \item We propose a two-step neural decoder with a simple decoding procedure in the first-step, applicable for all stabilizer codes.
    \item We suggest an alternative approach on combining the non-neural and the neural decoder which can be incorporated in other neural network based decoders.
    \item Our proposed approaches seem to have significant advantages with respect to training cost and complexity of the network for higher lengths when compared to the previous work of Maskara {\em et al.} \cite{maskara2018advantages}.
\end{compactenum}
The paper is organized as follows. We review the necessary background on Quantum Error Correction (QEC), ML and DL in Section~\ref{sec:Background}. We then describe our approach, the neural architecture used in detail and compare it with related work in Section~\ref{sec:OurWork}. In Section~\ref{sec:Insights}, we point out valuable insights from our work and conclude in Section~\ref{sec:Conclusion}.

\section{\label{sec:Background} Background}
In this section, we summarize the necessary background on Quantum Error Correcting Codes (QECC).
In Section~\ref{subsec:StabilizerFormalism}, we briefly review stabilizer codes.
In this paper we focus on color codes which are introduced in Section~\ref{subsec:ColorCodes}. Lastly, in the Section~\ref{subsec:Background:ML_DL} we describe basics of ML and DL with emphasis on deep learning by discussing the various components in a neural network which can be changed depending on the problem to be solved.

\subsection{\label{subsec:StabilizerFormalism} Stabilizer codes}
In this section, we  briefly review stabilizer codes.
Recall, that the Pauli group on a single qubit is generated by the Pauli matrices $\{\pm i I,X,Y,Z\}$.
The group $\mathcal{P}_n$ consists of tensor products on $n$ single qubit Pauli operators, $P_{1} \otimes P_{2}\otimes  ... \otimes  P_{n}$.
A stabilizer code is defined by an abelian subgroup $\mathcal{S} \subset \mathcal{P}_n$, such that $-I\not\in \mathcal{S}$.
The codespace $\mathcal{Q}$, is joint +1-eigenspace of $\mathcal{S}$. 
\begin{eqnarray*}
\mathcal{Q} = \{\: \ket \psi \in (\mathbb{C}^2)^{\otimes n} \mid S \ket \psi  = \ket \psi \: \text{ for all }\: S\in \mathcal{S} \: \}
\end{eqnarray*}

An $[[n,k]]$ stabilizer code   encodes $k$ logical qubits into $n$ physical qubits and its stabilizer $\mathcal{S}$ will have $n-k$ independent generators.
We assume that $\mathcal{S}$ is generated by $\mathcal{S}_g = \{S_1,\hdots, S_m \}$, where $m \geq n-k$ and  $S_1,\ldots, S_{n-k}$ are linearly independent.

Let $\mathcal{C}(S)$ be the centralizer of  $\mathcal{S}$ i.e. the set of all Pauli operators that commute with all the elements of $\mathcal{S}$.
Let   $\mathcal{L}_{g} =\{\overline{X}_{i},\overline{Z}_{i}\}_{i=1}^k$, where $\overline{X}_i$ and $\overline{Z}_{j}$ denote the logical $X$ and $Z$ operators of the code. 
Also, $\overline{X}_{i},\overline{Z}_{j}$ commute if $i \ne j$ and anti-commute if $i = j$.
Let $\mathcal{L} = \langle \overline{X}_{1}, \ldots, \overline{X}_{k}, \overline{Z}_{1}, \ldots, \overline{Z}_{k}\rangle$.

We define another set of operators $\mathcal{T}_{g}=\{ T_{1}, T_{2}, \hdots T_{n-k}\}$ called the pure errors,
such that $T_{i}$ and $S_{j}$ commute if $i \neq j$ and anti-commute if $i = j$.
The pure errors commute with each other and also with the logical operators. 
Let $\mathcal{T} = \langle T_1,\ldots, T_{n-k}\rangle$.
Note that $\{\mathcal{S}_{g}, \mathcal{L}_{g}, \mathcal{T}_{g}\}$ together form a generating set for $\mathcal{P}_{n}$. 

An error operator, $E \notin \mathcal{C}(S)$ will anti-commute with at least one stabilizer operator in group $\mathcal{S}$. If $E$ anti-commutes with the $i^{th}$ stabilizer $S_{i} \in \mathcal{S}$, the $i^{th}$ syndrome bit $s_i$ is one and zero otherwise.
By calculating the syndrome values for all the stabilizer generators, the syndrome vector can be written as, $\mathbf{s}=(s_{1}, s_{2}, ... , s_{m})$ where $m \geq n-k$.

We can write the error operator $E = T L S$ up to a phase as proposed in \cite{duclos2010renormalization}. Here $T \in \mathcal{T}$, $S \in \mathcal{S}$ and $L \in \mathcal{L}$. Note that the operators $T$, $L$, $S$ depend on the error $E$.
The effect of $S$ is trivial, implying two error patterns $E$ and $E'=\mathcal{S}E$ will have same effect on codespace. $\mathcal{S}$ introduces an equivalence relation in error operators and hence finding $S$ is of little interest. Also, given syndrome $\left(\mathbf{s}\right)$, we can uniquely identify $T$ but identifying $L$ is a difficult task.
The problem of error correction for stabilizer codes is finding the most likely $L$ given the syndrome vector $\mathbf{s}$.
Mathematically, we can write this as,
\begin{gather}
\widehat{L} = \underset{\gamma \> \in \> \mathcal{L}}{\argmax} \> Pr\left(\gamma \> \vert \> \mathbf{s}\right) = \underset{\gamma \> \in \> \mathcal{L}}{\argmax} \> \underset{\delta \> \in \> \mathcal{S}}{\sum} Pr\left(\gamma\delta \> \vert \> \mathbf{s}\right)
\label{eq:mlestimate}
\end{gather}
Decoding can be thought of as a classification problem.
We have $4^k$ classes, which is exponential in $k$ and this reformulation of the decoding problem as a classification is not much help for large $k$. 
Fortunately, surface codes and color codes have fixed number of logical operators for any length and this reformulation can be taken advantage of. 
However, this is not sufficient, note that the  computation of the probabilities in Eq.~\eqref{eq:mlestimate}, requires the summation over $2^{n-k}$ terms which is of exponential complexity. 
So the reformulation of the decoding as a classification is not adequate, but further work is required to fully exploit this perspective. 

\subsection{\label{subsec:ColorCodes} Color codes}
Topological codes are a class of stabilizer codes where the stabilizer generators are spatially local. Popular examples of topological codes are Toric codes \cite{KITAEV20032} and Color codes \cite{bombin06}.
Color codes are defined using a lattice embedded on a surface. 
Every vertex is trivalent and faces are 3-colorable.

Qubits are placed on the vertices of the lattice and for each face $f$, we define an $X$ and $Z$ type operators called the face operators.
We define the the stabilizers as,
\begin{gather*}
Z^{\left(f\right)}  = \underset{v \in  f}{\prod} Z_{v}, \hspace{1cm}
X^{\left(f\right)}  = \underset{v \in f}{\prod} X_{v}
\end{gather*}

\begin{figure}[h]
\centering
\includegraphics[scale=0.8]{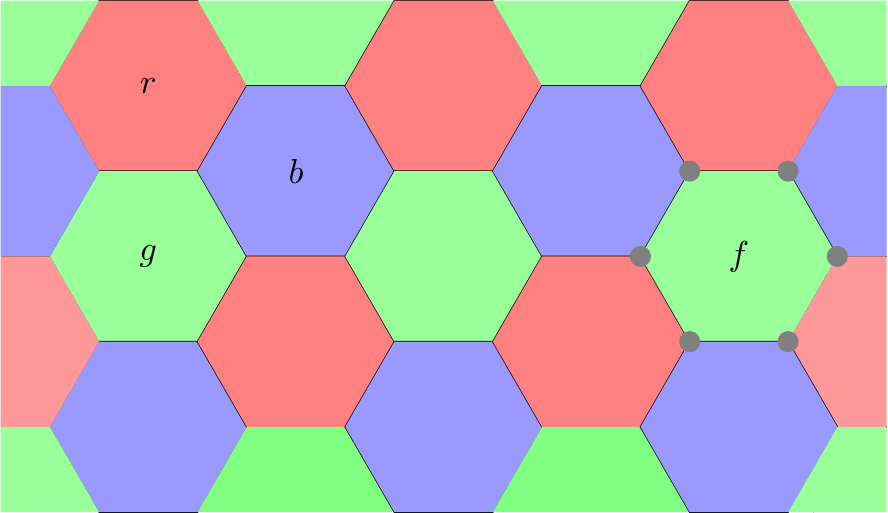}
\caption{\label{fig:colorCode}Periodic color code on a hexagonal lattice illustrated with a face and a stabilizer.}
\end{figure}

All $X$ and $Z$ type operators corresponding to every face generate the stabilizers of the color code. 
The color code on a hexagonal lattice with periodic boundary is shown in the Fig.~\ref{fig:colorCode}. It encodes four logical qubits~\cite{bombin06}.

\subsection{\label{subsec:Background:ML_DL}Machine Learning and Deep Learning}

\subsubsection{\label{subsubsec:ML}An overview of Machine Learning}
In traditional computing, algorithms are sets of explicitly programmed instructions which perform a specific task as to give out correct output for the given input.
ML is a concept to learn patterns from data through statistical analysis and make predictions without those rules being programmed explicitly. 
These ML algorithms are therefore data driven methods and the process of learning these rules or patterns is called training of the ML model. Training is essentially an optimization process minimizing an objective function called the loss function. This loss function plays an important role in the algorithm learning these patterns and making good predictions.

There are many such algorithms for solving problems of classification, regression etc and some of them are mentioned in~\cite{Kotsiantis:2007:SML:1566770.1566773,Domingos:2012:FUT:2347736.2347755}. Any function can be used as a loss function but they need not necessarily help the algorithm learn. There exist specific loss functions which are mathematically proven to be apt for solving each of the above mentioned tasks.
Mathematically, the core of any ML algorithm is to estimate the parameters of a function or set of functions which solve the given task.

Training can be classified into two types, supervised learning and the unsupervised learning. 
The requirement for supervised learning is labeled dataset of inputs $\left(\mathbf{x}\right)$ and the corresponding true outputs $\left(\mathbf{y}\right)$. These true outputs are sometimes referred to as ground truth. The ML algorithm will learn the patterns in the data by this information of input and correct output during training and tries to predict $\left(\mathbf{\widehat{y}}\right)$, the correct prediction during testing. Eg. Classification, Regression.
In unsupervised learning, we still have input data but the corresponding ground truth information is not present. The ML algorithm is required to learn the patterns from the input data alone without the information of the ground truth. Eg. Clustering.

\subsubsection{\label{subsubsec:DL}An overview of Deep Learning}
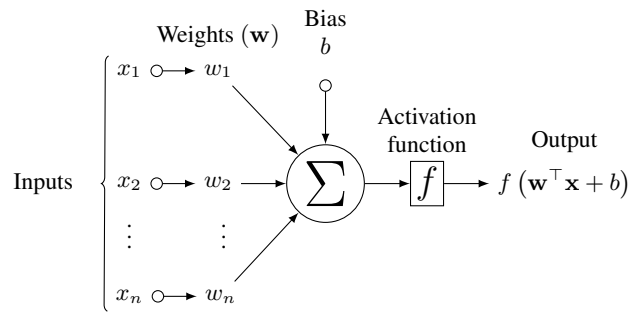
\begin{figure}
\centering
\begin{tikzpicture}[
init/.style={
  draw,
  circle,
  inner sep=2pt,
  font=\Huge,
  join = by -latex
},
squa/.style={
  draw,
  inner sep=2pt,
  font=\Large,
  join = by -latex
},
start chain=2,node distance=6mm
]
\node[on chain=2] 
  (x2) {$x_2$};
\node[on chain=2,join=by o-latex] 
  {$w_2$};
\node[on chain=2,init] (sigma) 
  {$\displaystyle\Sigma$};
\node[on chain=2,squa,label=above:{\parbox{2cm}{\centering Activation \\ function}}]   
  {$f$};
\node[on chain=2,label=above:Output,join=by -latex] 
  {$f\left(\mathbf{w}^{\top}\mathbf{x}+b\right)$};
\begin{scope}[start chain=1]
\node[on chain=1] at (0,1.5cm) 
  (x1) {$x_1$};
\node[on chain=1,label=above:{Weights $\left(\mathbf{w}\right)$},join=by o-latex] 
  (w1) {$w_1$};
  (w1) {$\vdots$};
\end{scope}
\begin{scope}[start chain=3]
\node[on chain=3] at (0,-1.5cm) 
  (x3) {$x_n$};
\node[on chain=3,join=by o-latex] 
  (w3) {$w_n$};
\end{scope}
\node[label=above:\parbox{2cm}{\centering Bias \\ $b$}] at (sigma|-w1) (b) {};
\begin{scope}[start chain=4]
\node[on chain=4] at (0,-0.6cm) 
  (x4) {$\vdots$};
\node[on chain=4] at (0.5cm, -0.6cm) 
  (w4) {$\vdots$};
\end{scope}

\draw[-latex] (w1) -- (sigma);
\draw[-latex] (w3) -- (sigma);
\draw[o-latex] (b) -- (sigma);

\draw[decorate,decoration={brace,mirror}] (x1.north west) -- node[left=10pt] {Inputs} (x3.south west);
\end{tikzpicture}
\caption{\label{fig:neuron} A single neuron which accepts input $\mathbf{x}$ and outputs $f\left(\mathbf{w}^{\top}\mathbf{x} + b\right)$ where $f$ is an activation function. The vectors $\mathbf{x}, \mathbf{w} \in \mathbb{R}^{n}$ and $b \in \mathbb{R}$.}
\end{figure}

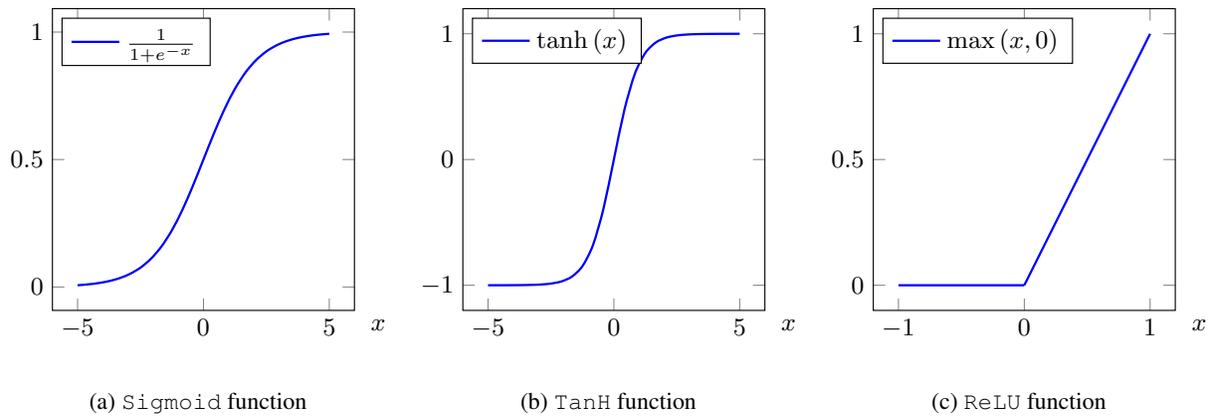
\begin{figure*}
\captionsetup[subfigure]{justification=centering}
\begin{subfigure}[b]{0.3\textwidth}
    \centering
    \resizebox{\linewidth}{!}{
    	\begin{tikzpicture}
        \begin{axis}[width=\textwidth, height=\textwidth, xlabel={$x$}, legend pos=north west, 
            every axis x label/.style={
                at={(ticklabel* cs:1.025)},
                anchor=north west
            }
        ]
        	\addplot[smooth, thick, solid, color=blue] {1/(1+exp{-x})};
      		\legend{$\frac{1}{1+e^{-x}}$}
        \end{axis}
      	\end{tikzpicture}
  	}
    \label{subfig:sigmoid}
    \subcaption{\texttt{Sigmoid} function}
\end{subfigure}
\begin{subfigure}[b]{0.3\textwidth}
    \centering
    \resizebox{\linewidth}{!}{
    	\begin{tikzpicture}
        \begin{axis}[width=\textwidth, height=\textwidth, xlabel={$x$}, legend pos=north west, 
            every axis x label/.style={
                at={(ticklabel* cs:1.025)},
                anchor=north west
            }
        ]
        	\addplot[smooth, thick, solid, color=blue] {tanh(x)};
      		\legend{$\tanh\left(x\right)$}
        \end{axis}
      	\end{tikzpicture}
  	}
    \label{subfig:tanh}
    \subcaption{\texttt{TanH} function}
\end{subfigure}
\begin{subfigure}[b]{0.3\textwidth}
    \centering
    \resizebox{\linewidth}{!}{
    	\begin{tikzpicture}
        \begin{axis}[width=\textwidth, height=\textwidth, xlabel={$x$}, legend pos=north west, domain=-3:3, 
            every axis x label/.style={
                at={(ticklabel* cs:1.025)},
                anchor=north west
            }
        ]
        	\addplot+[mark=none, smooth, thick, blue, domain=-1:0] {0};
            \addplot+[mark=none, smooth, thick, blue, domain=0:1] {x};
      		\legend{$\max\left(x,0\right)$\\}
        \end{axis}
      	\end{tikzpicture}
  	}
    \label{subfig:relu}
    \subcaption{\texttt{ReLU} function}
\end{subfigure}
\caption{\label{fig:activationFunctions} Various activation functions used commonly in DL. Note that \texttt{ReLU} does not saturate for high inputs.}
\end{figure*}

\noindent
\textbf{\textit{Neuron and Activation functions:}}
A \textit{neuron} is an element which takes an input $\mathbf{x}$ and performs the operation $f\left(\mathbf{w}^{\top}\mathbf{x} + b\right)$ as shown in the Fig.~\ref{fig:neuron}. 
The parameters $\mathbf{w}$ are called weights and the parameter $b$ is called the bias. 
Each element of these vectors $\mathbf{x}, \mathbf{w}$ and $b$ are real numbers. The function $f$ is a non-linear function and is called the \textit{activation function}. Some common activation functions include \verb+Sigmoid+, \verb+TanH+, \verb+ReLU+ (Rectified Linear Unit) etc as shown in the Fig.~\ref{fig:activationFunctions} and are exhaustively discussed in~\cite{Goodfellow-et-al-2016}.

Deep Learning (DL) is a method in ML to estimate the parameters of a function using a combinations of this basic element neuron.
It is common to address the combined set of parameters in $\mathbf{w}$ and $b$ as weights or parameters and we follow this same convention in our subsequent discussion.
The activation function plays a very crucial role in DL since without that, a neuron just performs a linear operation.

\noindent
\textbf{\textit{Architectures:}}
Different combinations of these basic neurons result in different architectures. 
Some of such famous architectures are Fully-Connected Networks (FC), Convolutional Neural Networks (CNN), Recurrent Neural Networks (RNN) etc. 
All these architectures comprise of layers which are again a combination of neurons. Essentially, these architectures can be characterized by these layers.

\noindent
\textbf{\textit{Fully-connected Network:}}
We briefly describe the FC architecture which we use in our work as shown in Fig.~\ref{fig:FCNetwork}. Any FC network has an input layer, an output layer and hidden layers. Each layer comprises of neurons and each neuron is connected to every other neuron in the adjacent layers. Connectedness implies that each neuron receives the output of the neurons it is connected to in the previous layer and it passes the output of itself to all the connected neurons in the next layer. All the neurons in every layer follow this rule except that the neurons in the input layer take the input from the data and the neurons in the output layer give us the final prediction. The input data and the output prediction varies from problem to problem. In a simple image classification task, the input data is the image and the output is the class label.
As mentioned before, the non-linear function plays a crucial role in the success of DL in estimating complicated functions efficiently, making DL a very powerful tool.
\tikzset{%
  every neuron/.style={
    circle,
    draw,
    minimum size=1cm
  },
  neuron missing/.style={
    draw=none, 
    scale=4,
    text height=0.333cm,
    execute at begin node=\color{black}$\vdots$
  },
}
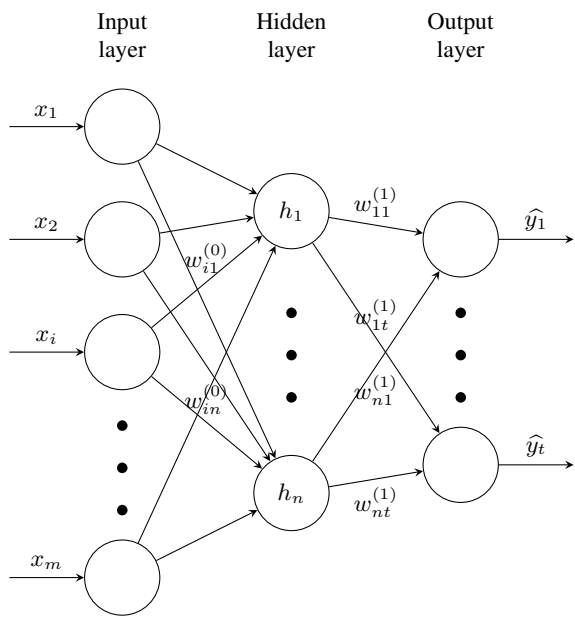
\begin{figure}
\centering
\begin{tikzpicture}[x=1.5cm, y=1.5cm, >=stealth]
\foreach \m/\l [count=\y] in {1,2,3,missing,4}
  \node [every neuron/.try, neuron \m/.try] (input-\m) at (0,2.5-\y) {};

\foreach \m [count=\y] in {1,missing,2}
  \node [every neuron/.try, neuron \m/.try ] (hidden-\m) at (1.5,2-\y*1.25) {};

\foreach \m [count=\y] in {1,missing,2}
  \node [every neuron/.try, neuron \m/.try ] (output-\m) at (3,1.5-\y) {};

\foreach \l [count=\i] in {1,2,i,m}
  \draw [<-] (input-\i) -- ++(-1,0)
    node [above, midway] {$x_\l$};

\foreach \l [count=\i] in {1,n}
  \node [] at (hidden-\i.center) {$h_\l$};

\foreach \l [count=\i] in {1,t}
  \draw [->] (output-\i) -- ++(1,0)
    node [above, midway] {$\widehat{y_\l}$};

\foreach \i in {1,...,2}
  \foreach \j in {1,...,2}
    \draw [->] (input-\i) -- (hidden-\j);
\foreach \i in {3}
  \foreach \j in {1}
    \draw [->] (input-\i) -- (hidden-\j)
    	node [above, midway] {$w^{(0)}_{i\j}$};
\foreach \i in {3}
  \foreach \j in {2}
    \draw [->] (input-\i) -- (hidden-\j)
    	node [above, midway] {$w^{(0)}_{in}$};
\foreach \i in {4}
  \foreach \j in {1,...,2}
    \draw [->] (input-\i) -- (hidden-\j);

\foreach \i in {1}
  \foreach \j in {1}
    \draw [->] (hidden-\i) -- (output-\j)
    	node [above, midway] {$w^{(1)}_{\i\j}$};
\foreach \i in {1}
  \foreach \j in {2}
    \draw [->] (hidden-\i) -- (output-\j)
    	node [above, midway] {$w^{(1)}_{\i t}$};
\foreach \i in {2}
  \foreach \j in {1}
    \draw [->] (hidden-\i) -- (output-\j)
    	node [below, midway] {$w^{(1)}_{n\j}$};
\foreach \i in {2}
  \foreach \j in {2}
    \draw [->] (hidden-\i) -- (output-\j)
    	node [below, midway] {$w^{(1)}_{nt}$};        

\foreach \l [count=\x from 0] in {Input, Hidden, Output}
  \node [align=center, above] at (\x*1.5,2) {\l \\ layer};
\end{tikzpicture}
\caption{\label{fig:FCNetwork} A sample fully-connected architecture with one hidden layer. Each neuron in every layer is connected to every other neuron in the adjacent layers. In this example, the size of the input vector is $m$ and the size of the output vector is $t$. There are $n$ hidden nodes in the hidden layer. The parameters $\mathbf{w}$ represent the weights of the network.}
\end{figure}

\noindent
\textbf{\textit{Loss functions:}}
The loss function plays a prominent role in the performance of any DL model. It is calculated between the true label $\left(\mathbf{y}\right)$ or the ground truth and the prediction made by the network $\left(\mathbf{\widehat{y}}\right)$. The training procedure as described next ensures that the predictions made by the network get closer to the ground truth by minimizing the loss function as the training progresses. For regression problem, commonly used loss functions are are $\ell_{2}$ and $\ell_{1}$ norms as defined below.
\begin{gather*}
\ell_{2}\left(\mathbf{y}, \mathbf{\widehat{y}}\right) = \left\lVert \mathbf{y} - \mathbf{\widehat{y}}\right\rVert_{2} = \sum_{i} \left(y_{i} - \widehat{y_{i}}\right)^{2} \\
\ell_{1}\left(\mathbf{y}, \mathbf{\widehat{y}}\right) = \left\lVert \mathbf{y} - \mathbf{\widehat{y}}\right\rVert_{1} = \sum_{i} \left\vert y_{i} - \widehat{y_{i}}\right\vert 
\end{gather*}
For classification problems, \textit{cross-entropy} ($\ell_{CE}$) is used as the loss function which is defined in the following equation. 
\begin{gather*}
\ell_{CE}\left(\mathbf{y}, \mathbf{\widehat{y}}\right) = - \sum_{i}y_{i}\log \left(\widehat{y}_{i}\right)
\end{gather*}
We use this cross-entropy loss in our work since QEC can be viewed as a classification problem as described in Section~\ref{subsec:OurWork:ProblemModeling}. We discuss the reasons for using this loss in Section~\ref{subsec:OurWork:TrainingProcedure}. 

\noindent
\textbf{\textit{Training:}}
Training is nothing but estimating the values of the weights of the network which minimizes the chosen loss function for the given training data or the input-output pairs. One of the traditional method of updating the weights to minimize a function is \textit{Gradient Descent} (GD) algorithm. It is an iterative algorithm which tries to optimize the objective function and in our case minimize the loss function $\left(\ell\right)$ through updating the weights $\left(\mathbf{w}\right)$ of the network in each iteration by following the update rule defined below, as discussed in~\cite{Goodfellow-et-al-2016}. 
\begin{gather*}
\mathbf{w}_{t+1} = \mathbf{w}_{t} - \alpha \nabla_{\mathbf{w}} \ell\left(\mathbf{y}, \mathbf{x}, \mathbf{w}_{t}\right)
\end{gather*}

Here, $\mathbf{w}_{i}$ are the weights of the network at the $i^{th}$ iteration. The weights $\mathbf{w}_{0}$ are initialized randomly. There are many methods to initialize these weights and we mention about them shortly. The parameter $\alpha$ is called the \textit{learning-rate} and is a \textit{hyper-parameter}. There are many such hyper-parameters and we also discuss them later in this section. The speed with which and the optima to which the model converges to, depends on $\alpha$. 

The gradient descent algorithm requires us to train on the entire training dataset at once, i.e calculate the average loss for all the inputs in the dataset and update the weights. Since that is not usually computationally feasible, a popular variant of it called the \textit{Stochastic Gradient Descent} (SGD) is employed. Instead of training on the entire dataset at once, the model is trained on small batches of data until all the training data is exhausted which completes one \textit{epoch}. The size of this batch is called the \textit{batch-size} as mentioned in~\cite{Goodfellow-et-al-2016}. For example, if the entire dataset contains $1000$ data points, then GD requires us to calculate the average loss on all the $1000$ inputs and then update the weights in one iteration. In SGD, say we choose the batch-size to be $50$, then $50$ data points are chosen randomly from the entire dataset of $1000$. The average loss is calculated for that batch of $50$ and the weights are updated. This completes one iteration. In the second iteration, another set of $50$ data points are chosen randomly from the remaining $950$ data points and the rest of the procedure follows. In this example, a total of $20$ iterations are required to exhaust the entire dataset which completes an epoch.

One of the major limitation of gradient descent and its variants is that it does not guarantee convergence to global optima. Since the loss is calculated between the true label $\left(\mathbf{y}\right)$ and the prediction of the network $\left(\mathbf{\widehat{y}}\right)$, it is indirectly a function of the weights of the network $\mathbf{w}$, since $\mathbf{\widehat{y}}$ is a function of $\mathbf{w}$ and $\mathbf{x}$.

\noindent
\textbf{\textit{Weight initialization and back-propagation:}}
Before training, the weights of the NN, $\mathbf{w}$ are randomly initialized. Weight initialization plays a crucial role in training and performance of the NN. There are many weight initialization methods but the popular ones are proposed by~\cite{He:2015:DDR:2919332.2919814} and~\cite{pmlr-v9-glorot10a}. These methods have been shown to perform well in solving classification problems. Training neural networks can be incredibly costly with GD or SGD but with the use of a dynamic programming based algorithm called the \textit{back-propagation} algorithm, the cost of training reduces significantly as discussed in~\cite{Goodfellow-et-al-2016}. The back-propagation algorithm also uses gradient-descent but stores the values of the gradients to the current layer in order to calculate the gradients to the weights of the previous layer.

\noindent
\textbf{\textit{Optimizers:}}
There are many variants of the SGD algorithm described above like RMSProp, AdaGrad as mentioned in~\cite{Goodfellow-et-al-2016} which have a modified update rule. All these rules are commonly called \textit{optimizers} since they optimize the weights of our network in order to minimize the loss function. We use \textit{Adam} optimizer, proposed by \cite{kingma2014adam} because of the significant improvements it offers during training and also in the performance of deep neural networks.

\noindent
\textbf{\textit{Hyper-parameters:}}
As we can see, numerous design decisions are required to build a neural network like the architecture, the loss function, activation function, weight initialization, optimizer etc. Once those are selected, we have few more parameters to experiment with, listed as follows,
\begin{compactenum}[i)]
  \item The number of hidden layers 
  \item The learning rate
  \item The number of neurons in each layer 
  \item The batch-size
\end{compactenum}
These parameters are called \textit{hyper-parameters} of the network. Choosing the right set of hyper-parameters for a give problem is one of the biggest challenges of DL. These parameters play a crucial role in both training and performance of the networks because the training procedure does not guarantee convergence to global minima of the loss function, as mentioned previously.

\subsubsection{\label{subsubsec:DL:process_flow} Process flow of a common DL architecture}
\begin{figure}
\centering
\captionsetup{justification=centering}
\includegraphics[scale=0.57]{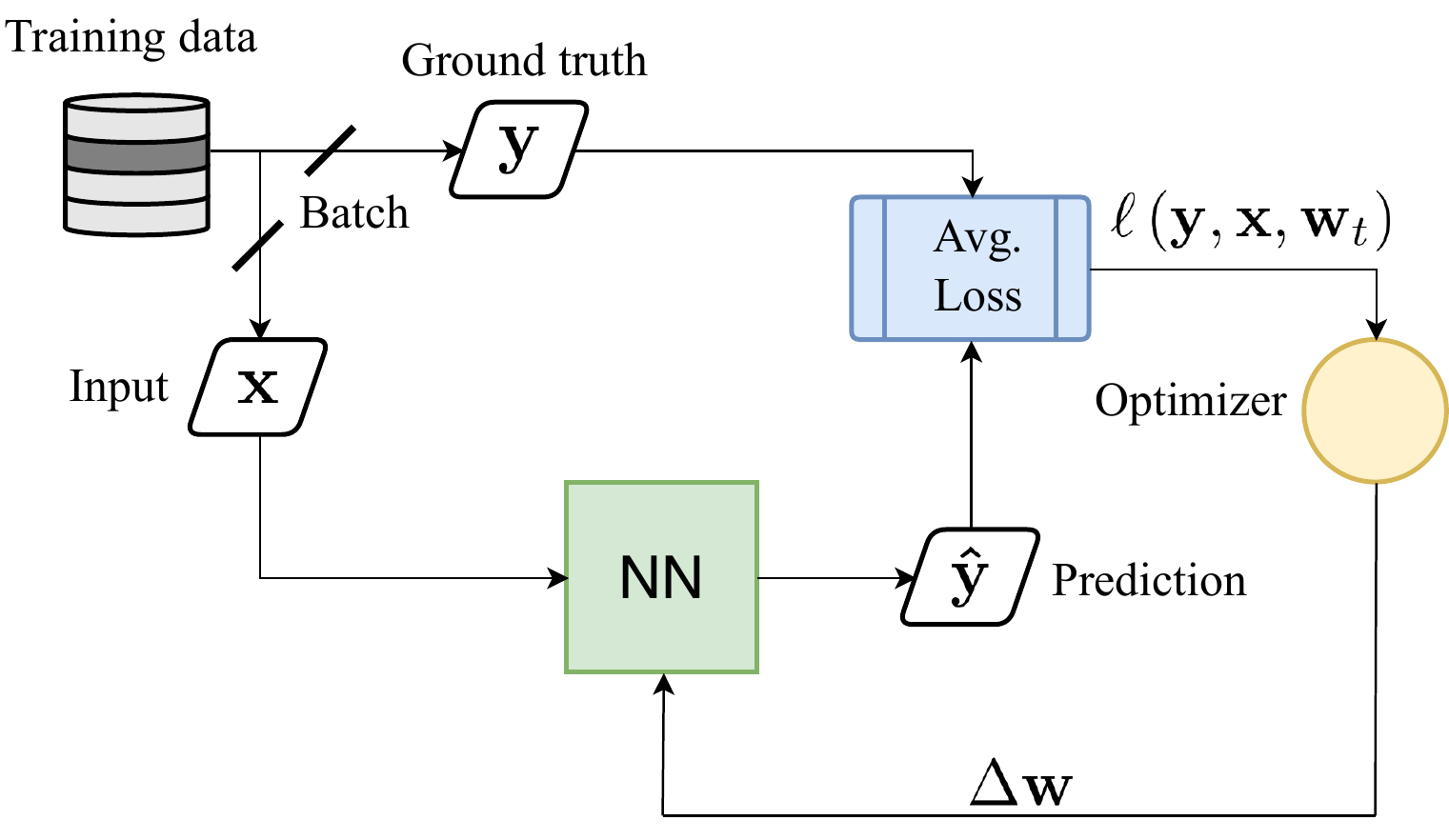}
\caption{\label{fig:DLProcessFlow}The process flow of any deep learning network. The NN represents any neural network either FC, CNN, RNN etc. It takes input $\mathbf{x}$ and makes the prediction $\mathbf{\widehat{y}}$. The loss is calculated between the ground truth $\mathbf{y}$ and the prediction $\mathbf{\widehat{y}}$ using the weights during the iteration $t$. The optimizer calculates the updates $\Delta \mathbf{w}$ according to the update rule and modifies the weights of the network for the $\left(t+1\right)^{th}$ iteration.}
\end{figure}

The process flow of any DL architecture can be modeled as shown in Fig.~\ref{fig:DLProcessFlow}. The NN can be any neural network as described previously. The NN takes an input $\mathbf{x}$ from the training data and makes a prediction $\mathbf{\widehat{y}}$. The loss is calculated between the ground truth $\mathbf{y}$ and the prediction $\mathbf{\widehat{y}}$. The optimizer then updates the weights of the NN according to the update rule. This whole process completes one iteration during training. We repeat this process until the loss  value between $\mathbf{y}$ and $\mathbf{\widehat{y}}$ saturates over multiple iterations.

\subsubsection{\label{subsubsec:ML:classification} Classification problem}
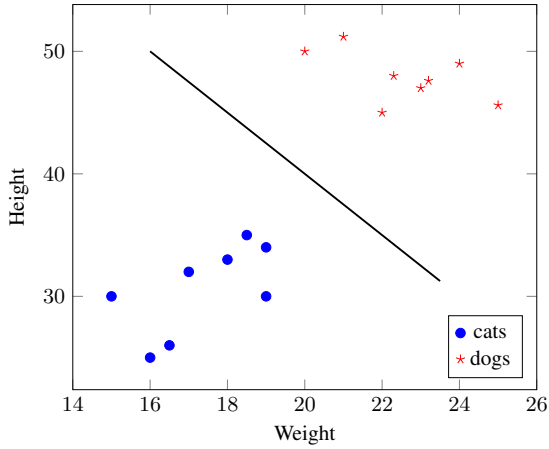
\begin{figure}
\centering
\begin{tikzpicture}[scale=0.9]
\begin{axis}[xlabel={Weight}, ylabel={Height}, legend pos=south east]
	\addplot[color=blue, only marks, mark=*] plot coordinates {    
    	(15, 30)
        (16, 25)
        (16.5, 26)
        (17, 32)
        (19, 30)
        (18.5, 35)
        (18, 33)
        (19, 34)
	};
	\addplot[color=red, only marks, mark=star] plot coordinates {    
    	(20, 50)
        (22, 45)
        (23, 47)
        (21, 51.2)
        (25, 45.6)
        (24, 49)
        (22.3, 48)
        (23.2, 47.6)
	};
	\legend{cats, dogs}
	\addplot+[mark=none, smooth, thick, black, domain=16:23.5] {90 - 2.5*x};
\end{axis}
\end{tikzpicture}
\caption{\label{fig:classification} Simple classification between domestic cats and dogs depending on weight and height using dummy data. Estimating the parameters of the boundary solves the classification problem.}
\end{figure}

In machine learning and statistics, classification is the problem of identifying to which of a set of categories or classes a new observation belongs to. This relation is statistically obtained from training data. A classification algorithm will predict the confidence score or the probability of the new observation belonging to a particular class. This can be illustrated in a dummy example of classification between domestic cats and dogs with the knowledge of their weight and length as shown in Fig.~\ref{fig:classification}. The weight and height are called the \textit{features} since the algorithm classifies with that information. Estimating the parameters of the line is solving the classification problem. In general the boundary could be a complicated curve and there could be multiple classes with multiple features. Commonly, these features might not be available and we have to devise algorithms to extract them from the input.

Mathematically, if we assume the feature vector to be $\mathbf{f}$ for an observation $x$ and the total classes are the set $\mathcal{C}$, then the prediction $\widehat{y}$ is the most likely class that $x$ belongs to as defined in the following equation.
\begin{gather*}
\widehat{y} = \underset{c \> \in \> \mathcal{C}}{\argmax} \> Pr\left(x \in c \> \vert \> \mathbf{f}\right) 
\end{gather*}

Generally, traditional ML algorithms requires us to extract these features $\left(\mathbf{f}\right)$ from the input $\left(\mathbf{x}\right)$ using some rules where as neural networks are known to extract them by themselves from the input directly, for example as shown in \cite{krizhevsky2012imagenet}. This helps immensely in the success of DL since the network learns to extract the important features for solving the problem, instead of us using hand coded rules to extract what we think are important features.

\section{\label{sec:OurWork} Decoding Color Codes using Neural Networks}
In this section, we describe our problem formulation for correction of phase errors and how the decoding can be modeled as a classification problem. For any stabilizer code, every error $E$ can be uniquely decomposed to the pure error $T$, logical error $L$ and a stabilizer $S$ as mentioned in the Section~\ref{subsec:StabilizerFormalism}. 
\begin{gather*}
E = T  L  S
\end{gather*}
Given the syndrome $\mathbf{s}$, we can uniquely identify $T$. Since the stabilizers $S$ form the equivalence class, the decoding problem comes down to correctly estimating $L$ given $\mathbf{s}$. 
In this work, we study CSS codes which have two types of stabilizers, $X$ and $Z$. They can be written in the matrix form as,
\begin{eqnarray*}
\mathbf{S} = \left[ 
\begin{array}{cc}
\mathbf{H}_{X} & 0 \\
0 & \mathbf{H}_{Z}
\end{array}
\right]
\end{eqnarray*}

Phase errors create $X$ non-zero syndromes and hence we consider only $X$ stabilizers from now on. The matrix $\mathbf{H}_{X}$ represents the $X$ stabilizers and $\mathbf{H}_{Z}$ represents the $Z$ stabilizers. For 2D color codes, $\mathbf{H}_{X} = \mathbf{H}_{Z}$ and in the subsequent equations, we use $\mathbf{H}$ instead of $\mathbf{H}_{X}$ for simplicity. Denote the binary representation of $E$ as $\mathbf{e} \in \mathbb{F}_2^{n}$.
Then we can calculate the corresponding syndrome as,
\begin{gather}
\mathbf{s}^{\top} = \mathbf{H} \mathbf{e}^{\top}
\label{eq:synPC}
\end{gather}

The matrix $\mathbf{H}$ is not full rank. In color code, $X$ stabilizers corresponding to faces have two dependencies as mentioned in~\cite{PhysRevLett.97.180501}. We remove those two dependent stabilizers from the $\mathbf{H}$ matrix, one stabilizer each corresponding to two different colors and denote it as $\mathbf{H}_{f}$ which is full rank. We calculate the right pseudo-inverse of $\mathbf{H}_{f}$ and denote it as $\mathbf{H}_{f}^{\dagger}$. 
\begin{gather}
\mathbf{H}_{f} \mathbf{H}_{f}^{\dagger} = \mathbf{I} \label{eq:pseudoInverse}
\end{gather}
The resultant syndrome which does not list the syndromes calculated by the removed dependent stabilizers is denoted by $\mathbf{s}_{f}$ as shown below. 
\begin{gather}
\mathbf{s}_{f}^{\top} = \mathbf{H}_{f} \mathbf{e}^{\top} \label{eq:syndromeCalc}
\end{gather}

\subsection{\label{subsec:OurWork:ProblemModeling} QEC as a classification problem}
Researchers have previously studied the perspective of quantum error correction as a classification problem using neural networks \cite{varsamopoulos2017decoding,maskara2018advantages,chamberland2018deep}.
As mentioned before, we model our decoder as a two-step process. The first-step is a simple inversion where we calculate an estimate $\widehat{E}$ of the actual error $E$ which has occurred. We first calculate the syndrome from Eq.~\eqref{eq:syndromeCalc} and then estimate $\mathbf{\widehat{e}} \in \mathbb{F}_{2}^{n}$, the binary representation of the operator $\widehat{E}$ as follows,
\begin{gather}
\mathbf{\widehat{e}}^{\top} = \mathbf{H}_{f}^{\dagger} \mathbf{s}_{f}^{\top}
\label{eq:errorEstimateHInv}
\end{gather}
Note that the syndrome of the estimate $\mathbf{\widehat{e}}$ will be same as the syndrome of $\mathbf{e}$. Hence, they have the same pure error $T$.
\begin{gather}
\mathbf{H}_{f} \mathbf{\widehat{e}}^{\top} = \mathbf{H}_{f} \mathbf{H}_{f}^{\dagger} \mathbf{s}_{f}^{\top} = \mathbf{s}_{f}^{\top} \nonumber\\
\Longrightarrow \mathbf{H} \mathbf{\widehat{e}}^{\top} = \mathbf{H} \mathbf{e}^{\top} = \mathbf{s}^{\top} \label{eq:sameSyn}
\end{gather}

This estimate $\mathbf{\widehat{e}}$ computed using Eq.~\eqref{eq:errorEstimateHInv} need not be same as $\mathbf{e}$. This is because there exist multiple errors with the same syndrome. We have chosen one solution by fixing $\mathbf{H}_{f}^{\dagger}$ which is calculated only once. This makes the first-step of the decoder simple. From Eq.~\eqref{eq:sameSyn}, we can conclude that the pure error is same in both $E$ and $\widehat{E}$ and we denote it by $T$. Applying this initial estimate $\widehat{E}$ onto the system might result in logical error. This can be concluded through the following equations.
\begin{gather}
E = T L S \hspace{0.5cm} \text{and} \hspace{0.5cm} \widehat{E} = T \widehat{L} \widehat{S} \nonumber \\ 
\Longrightarrow \widehat{E} E = T \widehat{L} \widehat{S} \> T L S = \left(\pm\right) L\widehat{L} S\widehat{S} \nonumber \\
\Longrightarrow \widehat{E} E = \left(\pm\right) \widetilde{L} \widetilde{S} \label{eq:correctionHomology}
\end{gather}

Here $\widetilde{L} = L\widehat{L}$ and $\widetilde{S} = S\widehat{S}$. The reason for occurrence of $\left(\pm\right)$ in Eq.~\eqref{eq:correctionHomology} is because the Pauli operators $T$, $\widehat{S}$ might commute or anti-commute. This is of little interest to us because we estimate the error up to a global phase.

The homology of $\widehat{E} E$ is same as the homology of $\widetilde{L}$ since $\widetilde{S}$ has a trivial homology.
If we can predict the resultant homology $\widetilde{L}$, we can get back to the trivial state and the decoding succeeds. Since the number of homologies are fixed, this is modeled in the second-step of our decoder as a classification problem using NN. The goal of the NN is to predict $\widetilde{L}$ given the syndrome $\mathbf{s}$. Our final error correction will be,
\begin{gather}
\widetilde{E} = \widetilde{L} \widehat{E}
\label{eq:finalErrorCorrection}
\end{gather}

If the NN properly predicts $\widetilde{L}$ this correction will restore the state up to a global phase which is evident through the following equations.
\begin{gather}
\widetilde{E} E = \widetilde{L} \widehat{E} E \nonumber \\
\Longrightarrow \widetilde{E} E = \left(\pm\right) \widetilde{L} \widetilde{L} \widetilde{S} \nonumber \\
\Longrightarrow \widetilde{E} E = \left(\pm\right) \widetilde{S} \nonumber
\end{gather}

\begin{figure*}
\includegraphics[scale=0.55]{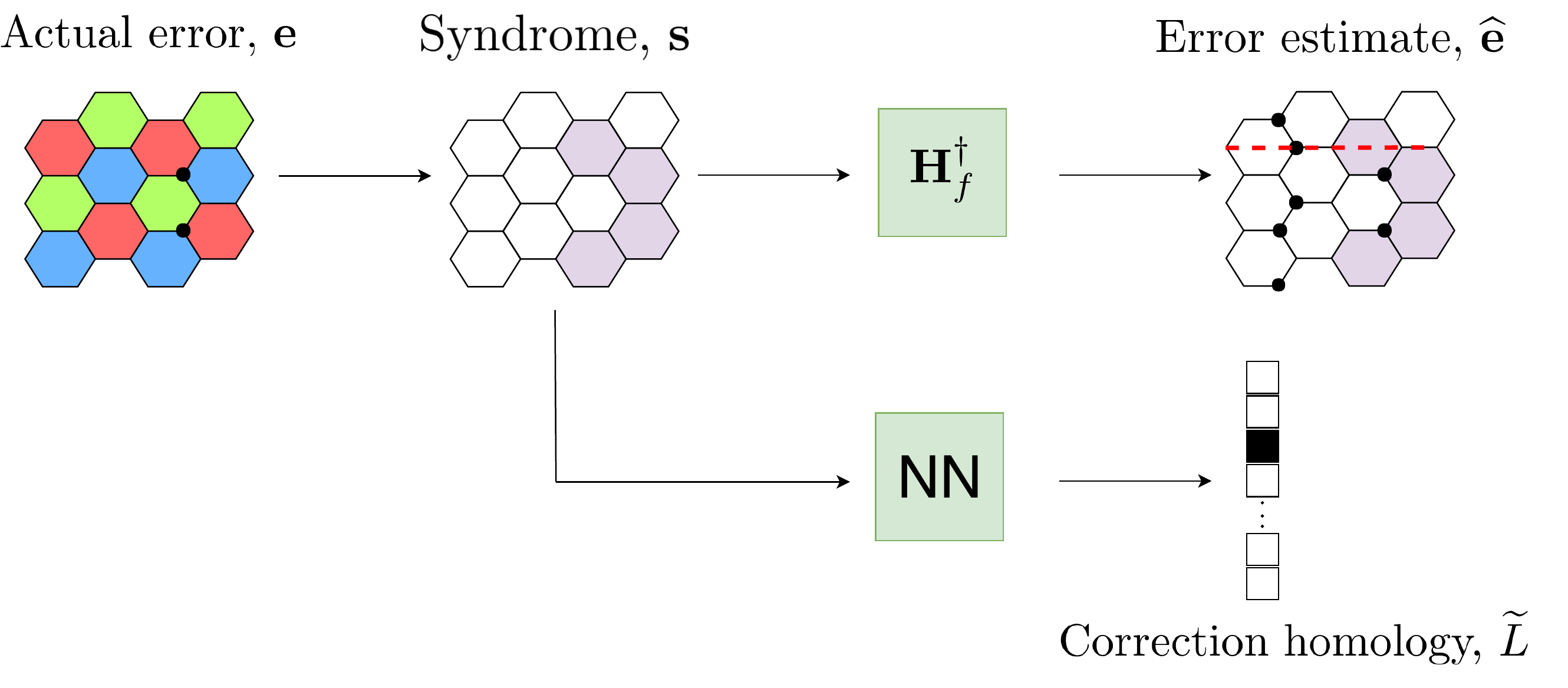}
\caption{\label{fig:decoderFlowDigOne} Flow diagram of our two-step decoder. The black dots represent error on the qubits and the marked regions represent the syndrome caused. In the first-step we get an estimate of the error $\mathbf{\widehat{e}}$ and in the second-step, we predict the correction homology $\widetilde{L}$ using our trained NN. Our final error correction is $\widetilde{L} \widehat{E}$. Refer Eqs.~\eqref{eq:errorEstimateHInv},~\eqref{eq:correctionHomology}, and~\eqref{eq:finalErrorCorrection}. Note that the $\mathbf{H}$-inverse decoder in step-one need not always give us pure error. In this example, the error estimate operator $\widehat{E}$ anti-commutes with a logical operator (red dashed line) and hence cannot be a pure error.}
\end{figure*}

The work by \cite{maskara2018advantages} used a naive decoder which removes syndromes by pushing errors to the boundary in the first-step. 
Their neural network tries to improve upon this estimate by predicting the correction homology. 
Mathematically, this means that their decoder could implement different inverse for a different syndrome. 
In our approach, we fix the inverse in the first-step, making our initial decoder much simpler. We discuss more on this in the Section~\ref{sec:Insights}.
The first-step decoder in \cite{varsamopoulos2017decoding} is to estimate the pure-error which needs to satisfy many properties.
We want to emphasize that our inverse matrix $\mathbf{H}_{f}^{\dagger}$ in step-one gives us an error estimate which need not always be pure error. It entirely depends on the construction of $\mathbf{H}_{f}^{\dagger}$. We used \textit{SageMath} \footnote{\url{http://sagemath.org}}, an open-source mathematics software for calculating $\mathbf{H}_{f}^{\dagger}$ from Eq.~\eqref{eq:pseudoInverse}.

\subsection{\label{subsec:OurWork:Architecture} Neural decoder}
In this section, we describe our neural decoder in the second-step. As mentioned before, we have modeled our NN in two ways and in both of them we have used a fully-connected architecture where every neuron in one layer is connected to every other neuron in the adjacent layers. The output of the network is the homology vector where each element of it represents a homology class. Since this is a classification problem, we use cross-entropy as our loss function which needs to be minimized during training. We have used Adam optimizer proposed by~\cite{kingma2014adam} since it has been observed to perform better than  the other optimizers in terms of convergence of the loss. We have also used 1D batch normalization layer after every layer in the network. It is proven to significantly boost the training speed as shown in~\cite{ioffe2015batch}. The activation function used for every neuron is \verb+ReLU+ since it has shown to perform well when compared to other functions like \verb+Sigmoid+ or \verb+TanH+ by reducing the problem of vanishing gradients as the network goes deeper as shown in~\cite{karlik2011performance,pmlr-v15-glorot11a}.

\begin{table}
\caption{\label{tab:hyperParameters}The values of the hyper-parameters used in the neural decoder in our first approach.}
\begin{ruledtabular}
\begin{tabular}{c|cccccc}
\diagbox{$d\footnote{Distance of the code}$}{parameters} & $h_{d}\footnote{Number of hidden layers}$ & $f_{d}\footnote{Hidden dimension factor}$ & $b_{d}\footnote{Batch size}$ & $\alpha\footnote{Learning rate}$ & $t_{d, p_{err}}\footnote{Number of training samples per each $p_{err}$}$ & $T_{d}\footnote{Total number of training samples for all $p_{err}$ combined}$\\
\hline
$6$ & $2$ & $2$ & $500$ & $0.001$ & $2 \times 10^{7}$ & $1.4 \times 10^{8}$\\
$8$ & $3$ & $5$ & $750$ & $0.001$ & $4 \times 10^{7}$ & $2.8 \times 10^{8}$\\
$9$ & $4$ & $5$ & $750$ & $0.001$ & $4 \times 10^{7}$ & $2.8 \times 10^{8}$\\
$12$ & $7$ & $10$ & $2500$ & $0.001$ & $10 \times 10^{7}$ & $7 \times 10^{8}$\\
\end{tabular}
\end{ruledtabular}
\end{table}

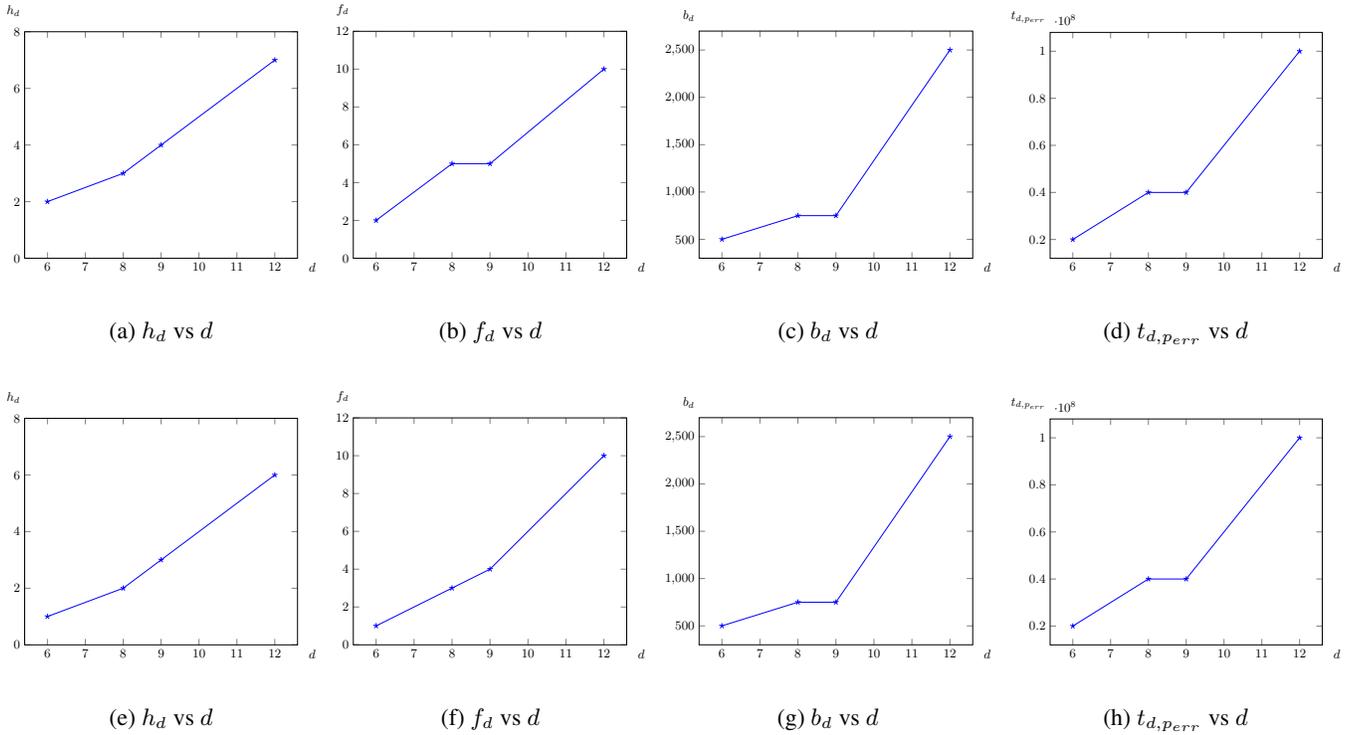
\begin{figure*}
\captionsetup[subfigure]{justification=centering}
\begin{subfigure}[b]{0.24\textwidth}
    \centering
    \resizebox{\linewidth}{!}{
    	\begin{tikzpicture}
          \begin{axis}[xlabel={$d$}, ylabel={$h_{d}$}, ymin=0, ymax=8, 
                every axis x label/.style={
                    at={(ticklabel* cs:1.025)},
                    anchor=north west
                },
                every axis y label/.style={
                    at={(ticklabel* cs:1.05)},
                    anchor=south east
                }
            ]
              \addplot[solid, color=blue, mark=star] plot coordinates {
                  (6, 2)
                  (8, 3)
                  (9, 4)
                  (12, 7)
              };
          \end{axis}    
      	\end{tikzpicture}
  	}
    \label{subfig:hdvsd}
    \subcaption{$h_{d}$ vs $d$}
\end{subfigure}
\begin{subfigure}[b]{0.24\textwidth}
    \centering
    \resizebox{\linewidth}{!}{
    	\begin{tikzpicture}
    		\begin{axis}[xlabel={$d$}, ylabel={$f_{d}$}, ymin=0, ymax=12, 
                every axis x label/.style={
                    at={(ticklabel* cs:1.025)},
                    anchor=north west
                },
                every axis y label/.style={
                    at={(ticklabel* cs:1.05)},
                    anchor=south east
                }
            ]
              \addplot[solid, color=blue, mark=star] plot coordinates {
                  (6, 2)
                  (8, 5)
                  (9, 5)
                  (12, 10)
              };
          \end{axis}    
      	\end{tikzpicture}
  	}
    \label{subfig:fdvsd}
    \subcaption{$f_{d}$ vs $d$}
\end{subfigure}
\begin{subfigure}[b]{0.2525\textwidth}
    \centering
    \resizebox{\linewidth}{!}{
    	\begin{tikzpicture}
    		\begin{axis}[xlabel={$d$}, ylabel={$b_{d}$}, 
                every axis x label/.style={
                    at={(ticklabel* cs:1.025)},
                    anchor=north west
                },
                every axis y label/.style={
                    at={(ticklabel* cs:1.025)},
                    anchor=south east
                }
            ]
              \addplot[solid, color=blue, mark=star] plot coordinates {
                  (6, 500)
                  (8, 750)
                  (9, 750)
                  (12, 2500)
              };
          \end{axis}    
      	\end{tikzpicture}
  	}
    \label{subfig:bdvsd}
    \subcaption{$b_{d}$ vs $d$}
\end{subfigure}
\begin{subfigure}[b]{0.255\textwidth}
    \centering
    \resizebox{\linewidth}{!}{
    	\begin{tikzpicture}
    		\begin{axis}[xlabel={$d$}, ylabel={$t_{d, p_{err}}$}, 
                every axis x label/.style={
                    at={(ticklabel* cs:1.025)},
                    anchor=north west
                },
                every axis y label/.style={
                    at={(ticklabel* cs:1.025)},
                    anchor=south east
                }
            ]
              \addplot[solid, color=blue, mark=star] plot coordinates {
                  (6, 2*10^7)
                  (8, 4*10^7)
                  (9, 4*10^7)
                  (12, 10*10^7)
              };
          \end{axis}    
      	\end{tikzpicture}
  	}
    \label{subfig:tdperrvsd}
    \subcaption{$t_{d, p_{err}}$ vs $d$}
\end{subfigure}

\bigskip

\begin{subfigure}[b]{0.24\textwidth}
    \centering
    \resizebox{\linewidth}{!}{
    	\begin{tikzpicture}
          \begin{axis}[xlabel={$d$}, ylabel={$h_{d}$}, ymin=0, ymax=8, 
                every axis x label/.style={
                    at={(ticklabel* cs:1.025)},
                    anchor=north west
                },
                every axis y label/.style={
                    at={(ticklabel* cs:1.05)},
                    anchor=south east
                }
            ]
              \addplot[solid, color=blue, mark=star] plot coordinates {
                  (6, 1)
                  (8, 2)
                  (9, 3)
                  (12, 6)
              };
          \end{axis}    
      	\end{tikzpicture}
  	}
    \label{subfig:hdvsd_two}
    \subcaption{$h_{d}$ vs $d$}
\end{subfigure}
\begin{subfigure}[b]{0.24\textwidth}
    \centering
    \resizebox{\linewidth}{!}{
    	\begin{tikzpicture}
    		\begin{axis}[xlabel={$d$}, ylabel={$f_{d}$}, ymin=0, ymax=12, 
                every axis x label/.style={
                    at={(ticklabel* cs:1.025)},
                    anchor=north west
                },
                every axis y label/.style={
                    at={(ticklabel* cs:1.05)},
                    anchor=south east
                }
            ]
              \addplot[solid, color=blue, mark=star] plot coordinates {
                  (6, 1)
                  (8, 3)
                  (9, 4)
                  (12, 10)
              };
          \end{axis}    
      	\end{tikzpicture}
  	}
    \label{subfig:fdvsd_two}
    \subcaption{$f_{d}$ vs $d$}
\end{subfigure}
\begin{subfigure}[b]{0.2525\textwidth}
    \centering
    \resizebox{\linewidth}{!}{
    	\begin{tikzpicture}
    		\begin{axis}[xlabel={$d$}, ylabel={$b_{d}$}, 
                every axis x label/.style={
                    at={(ticklabel* cs:1.025)},
                    anchor=north west
                },
                every axis y label/.style={
                    at={(ticklabel* cs:1.025)},
                    anchor=south east
                }
            ]
              \addplot[solid, color=blue, mark=star] plot coordinates {
                  (6, 500)
                  (8, 750)
                  (9, 750)
                  (12, 2500)
              };
          \end{axis}    
      	\end{tikzpicture}
  	}
    \label{subfig:bdvsd_two}
    \subcaption{$b_{d}$ vs $d$}
\end{subfigure}
\begin{subfigure}[b]{0.255\textwidth}
    \centering
    \resizebox{\linewidth}{!}{
    	\begin{tikzpicture}
    		\begin{axis}[xlabel={$d$}, ylabel={$t_{d, p_{err}}$}, 
                every axis x label/.style={
                    at={(ticklabel* cs:1.025)},
                    anchor=north west
                },
                every axis y label/.style={
                    at={(ticklabel* cs:1.025)},
                    anchor=south east
                }
            ]
              \addplot[solid, color=blue, mark=star] plot coordinates {
                  (6, 2*10^7)
                  (8, 4*10^7)
                  (9, 4*10^7)
                  (12, 10*10^7)
              };
          \end{axis}    
      	\end{tikzpicture}
  	}
    \label{subfig:tdperrvsd_two}
    \subcaption{$t_{d, p_{err}}$ vs $d$}
\end{subfigure}
\caption{\label{fig:HPvsD} Plots of the various hyper-parameters of our neural networks with the distance $d$ of the code. Figs.~(a)-(d) are for the first-approach and the Figs.~(e)-(h) are for the second-approach.}
\end{figure*}

\subsection{\label{subsec:OurWork:TrainingProcedure} Training procedure}
For the network to decode correctly, it needs to be trained. We employ a supervised training procedure where we have labeled data of input (we generate $\mathbf{e}$ according to the noise and calculate syndromes $\mathbf{s}$ from Eq.~\eqref{eq:synPC}) and the corresponding output (homology $\widetilde{L}$). This output is the ground truth. Training is nothing but an optimization process where the weights of the network are optimized to minimize an objective function. This objective function is called loss function. The loss function plays a crucial role during training since certain loss functions are apt for certain problems. Since our NN needs to solve a classification problem, we use cross-entropy as our loss function. This is because given a syndrome $\left(\mathbf{s}\right)$, the NN predicts a probability distribution over all the possible classes. If we assume input is $\mathbf{x}$, the output of the NN is a distribution $\mathbf{q}\left(\mathbf{x}\right)$ and the true distribution is $\mathbf{p}\left(\mathbf{x}\right)$, cross-entropy can be written as follows.
\begin{gather}
\ell_{CE}\left(\mathbf{p}, \mathbf{q}\right) = - \sum_{x}\mathbf{p}\left(\mathbf{x}\right)\log \mathbf{q}\left(\mathbf{x}\right)
\label{eq:crossEntropy}
\end{gather}
This is same as minimizing the Kullback-Liebler divergence $\left(D_{KL}\right)$ between the distributions $\mathbf{p}\left(\mathbf{x}\right)$ and $\mathbf{q}\left(\mathbf{x}\right)$ up to a constant since $D_{KL}\left(\mathbf{p} \Vert \mathbf{q}\right)$ can be written as,
\begin{gather}
D_{KL}\left(\mathbf{p} \Vert \mathbf{q}\right) = \ell_{CE}\left(\mathbf{p}, \mathbf{q}\right) - \sum_{x}\mathbf{p}\left(\mathbf{x}\right)\log \mathbf{p}\left(\mathbf{x}\right)
\nonumber
\end{gather}
and the term $\sum_{x}\mathbf{p}\left(\mathbf{x}\right)\log \mathbf{p}\left(\mathbf{x}\right)$ is a constant because it is completely determined by the true distribution $\mathbf{p}$. This implies that minimizing $\ell_{CE}$ in Eq.~\eqref{eq:crossEntropy} gets the distribution learned by our NN i.e, $\mathbf{q}$ closer to the true distribution $\mathbf{p}$. 

Given a syndrome vector $\mathbf{s}$, a trained NN should be able to correctly predict the correct correction homology class $\widetilde{L}$ for all error rates under the threshold. In order to train a NN which is independent of the error rate, we employ a progressive training procedure as described in \cite{maskara2018advantages}. We generate training samples at a fixed error rate $p_{err}$ in each case and we train our NN for that noise until the loss function in Eq.~\eqref{eq:crossEntropy} saturates. We then move on to a higher $p_{err}$ and repeat the process for various error rates under the threshold. For our experiments (bit-flip noise), we have trained our NN for the error rates $\left\{ 0.05, 0.06, 0.07, 0.08, 0.09, 0.10, 0.11 \right\}$. We use Xavier normal initialization for the parameters in fully-connected layers and Gaussian normal initialization for the parameters in batch-normalization layer before we start training. We do not reinitialize the weights during the progressive training while we train on the higher $p_{err}$. We discuss the importance of this progressive training with evidence in the Section~\ref{sec:Insights}. 
In our first approach, we use the syndrome $\mathbf{s}$ alone as the input to the network whereas in our second approach, we use the concatenated vector of both initial estimate $\widehat{\mathbf{e}}$ and the syndrome $\mathbf{s}$. In both cases, the network is trained to predict correction homology $\widetilde{L}$.
Our $\mathbf{H}$-inverse decoder in step-one can be summarized in Alg.~\ref{alg:HInvDec}. The neural decoders can be summarized in Algs.~\ref{alg:neuralDecFirst},~\ref{alg:neuralDecSecond} for our first and second approaches respectively. The architectures for our decoders are illustrated in Figs.~\ref{fig:decoderFlowDigOne},~\ref{fig:decoderFlowDigTwo} for first and second approaches respectively.

\begin{algorithm}[H]
\caption{$\mathbf{H}$-inverse decoder (step-one)}
\begin{algorithmic}[1]
    \REQUIRE {Syndrome vector $\mathbf{s}$ and requires pre-computed $\mathbf{H}_{f}^{\dagger}$ matrix}
    \ENSURE {Error estimate operator $\widehat{E}$}
    \STATE Compute $\mathbf{s}_{f}$ from $\mathbf{s}$ by removing the syndromes of the removed dependent stabilizers while computing the matrix $\mathbf{H}_{f}$
    \STATE Compute $\widehat{\mathbf{e}}^{\top} = \mathbf{H}_{f}^{\dagger}\mathbf{s}_{f}^{\top}$ \COMMENT{from Eq.~\eqref{eq:errorEstimateHInv}}
    \STATE Return $\widehat{E}$, the error operator of $\widehat{\mathbf{e}}$ as the initial error estimate
\end{algorithmic}
\label{alg:HInvDec}
\end{algorithm}

\begin{algorithm}[H]
\caption{Neural decoder (step-two, first approach)}
\begin{algorithmic}[1]
    \REQUIRE {Syndrome vector $\mathbf{s}$, requires the trained neural network to predict the correction homology $\widetilde{L}$ and the initial estimate $\widehat{E}$}
    \ENSURE {Final error correction operator $\widetilde{E}$}
    \STATE Using the trained neural network, predict the correction homology $\widetilde{L}$ by giving the syndrome vector $\mathbf{s}$ as the input 
    \STATE Compute $\widetilde{E} = \widetilde{L} \widehat{E}$ \COMMENT{from Eq.~\eqref{eq:finalErrorCorrection}}
    \STATE Return $\widetilde{E}$ as the final error correction
\end{algorithmic}
\label{alg:neuralDecFirst}
\end{algorithm}

\begin{algorithm}[H]
\caption{Neural decoder (step-two, second approach)}
\begin{algorithmic}[1]
    \REQUIRE {Syndrome vector $\mathbf{s}$ and the initial estimate $\widehat{E}$, requires the trained neural network to predict the correction homology $\widetilde{L}$}
    \ENSURE {Final error correction operator $\widetilde{E}$}
    \STATE Using the trained neural network, predict the correction homology $\widetilde{L}$ by giving the concatenated vector of initial estimate $\mathbf{\widehat{e}}$ and the syndrome $\mathbf{s}$ as the input 
    \STATE Compute $\widetilde{E} = \widetilde{L} \widehat{E}$ \COMMENT{from Eq.~\eqref{eq:finalErrorCorrection}}
    \STATE Return $\widetilde{E}$ as the final error correction
\end{algorithmic}
\label{alg:neuralDecSecond}
\end{algorithm}

\subsection{\label{subsec:OurWork:Results} Results}
\begin{figure}
\centering
\begin{tikzpicture}[scale=0.9]
\begin{semilogyaxis}[
    xlabel={$p_{err}$}, 
    ylabel={Logical error}, 
    legend pos=south east,
    every axis x label/.style={
        at={(ticklabel* cs:1.025)},
        anchor=north west
    },
    ymin=1e-1, ymax=1e0
]
	\addplot[solid, color=blue, mark=*] plot coordinates {
    	(0.01, 0.1956)
        (0.05, 0.6424)
        (0.10, 0.8418)
        (0.15, 0.9073)
        (0.20, 0.9275)
        (0.25, 0.9360)
	};
	\addplot[solid, color=red, mark=square*] plot coordinates {
    	(0.01, 0.2680)
        (0.05, 0.7513)
        (0.10, 0.8944)
        (0.15, 0.9269)
        (0.20, 0.9343)
        (0.25, 0.9365)
	};
	\addplot[solid, color=orange, mark=triangle*] plot coordinates {
    	(0.01, 0.2954)
        (0.05, 0.7817)
        (0.10, 0.9059)
        (0.15, 0.9303)
        (0.20, 0.9366)
        (0.25, 0.9395)
	};
	\addplot[solid, color=olive, mark=star] plot coordinates {
    	(0.01, 0.3585)
        (0.05, 0.8361)
        (0.10, 0.9211)
        (0.15, 0.9361)
        (0.20, 0.9364)
        (0.25, 0.9377)
	};
	\legend{$d=6$,$d=8$,$d=9$,$d=12$}
\end{semilogyaxis}
\end{tikzpicture}
\caption{\label{fig:HInvDecResults} Performance of our $\mathbf{H}$-inverse $\left(\mathbf{H}_{f}^{\dagger}\right)$ decoder in step-one. Note that it is a very bad decoder by itself since for a fixed $p_{err}$, the logical error increases as the length of the code increases and this decoder on its own does not have a threshold.}
\end{figure}
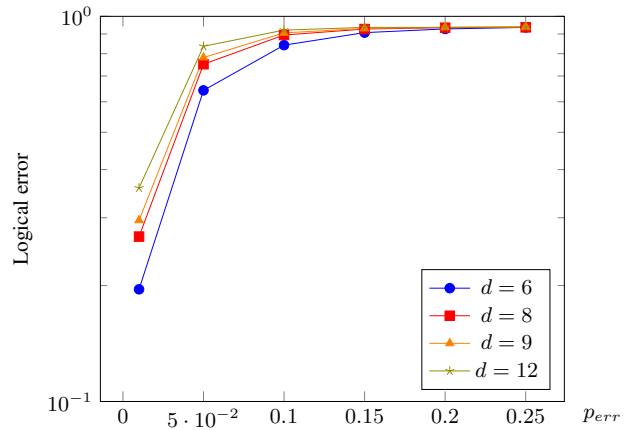

We describe our simulation results for bit-flip noise model in this section. As described earlier in the Section~\ref{sec:OurWork}, our decoder is a two-step decoder where we use a naive and deterministic $\mathbf{H}$-inverse $\left(\mathbf{H}_{f}^{\dagger}\right)$ decoder in step-one and then improve its performance in step-two using a NN. The performance of our $\mathbf{H}$-inverse decoder in the step-one by itself is shown in the Fig.~\ref{fig:HInvDecResults}. It shows that $\mathbf{H}$-inverse alone is a very bad decoder since the logical error increases as the length of the code increases for a fixed $p_{err}$. It is quite evident that this decoder does not have a threshold since the curves do not meet anywhere below the theoretical threshold of $10.97 \%$ \cite{PhysRevLett.103.090501}. 

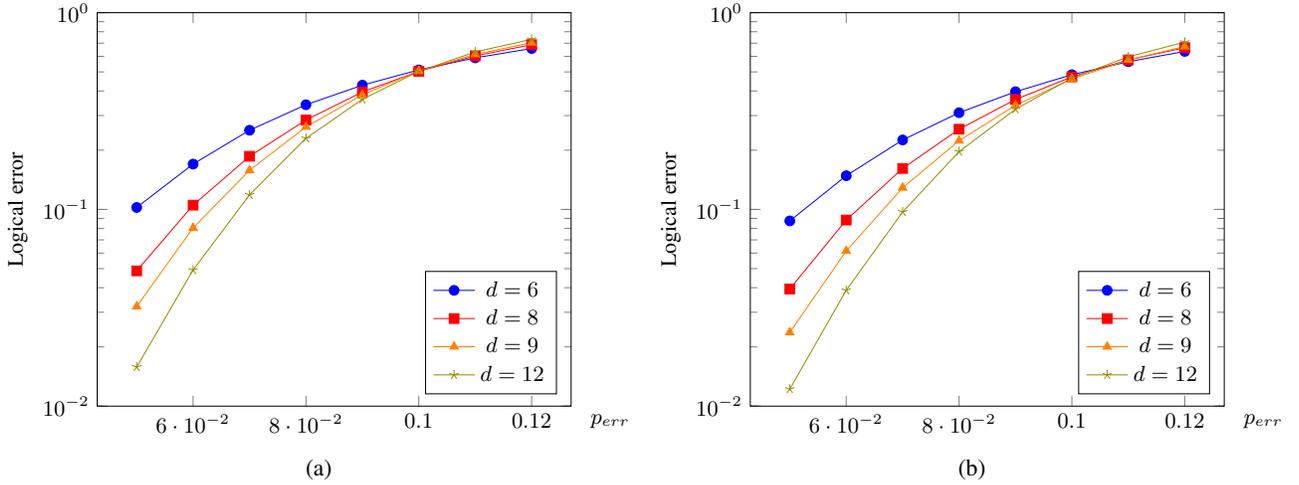
\begin{figure*}
\captionsetup[subfigure]{justification=centering}
\begin{subfigure}[b]{0.48\textwidth}
    \centering
    \resizebox{\linewidth}{!}{
    	\begin{tikzpicture}
            \begin{semilogyaxis}[
                xlabel={$p_{err}$}, 
                ylabel={Logical error}, 
                legend pos=south east,
                every axis x label/.style={
                    at={(ticklabel* cs:1.025)},
                    anchor=north west
                },
                ymin=1e-2, ymax=1e0
            ]
            	\addplot[solid, color=blue, mark=*] plot coordinates {    
                	(0.05, 0.1023)
                    (0.06, 0.1700)
                    (0.07, 0.2526)
                    (0.08, 0.3404)
                    (0.09, 0.4281)
                    (0.10, 0.5120)
                    (0.11, 0.5895)
                    (0.12, 0.6576)
            	};
            	\addplot[solid, color=red, mark=square*] plot coordinates {    
                	(0.05, 0.0487)
                    (0.06, 0.1051)
                    (0.07, 0.1862)
                    (0.08, 0.2847)
                    (0.09, 0.3954)
                    (0.10, 0.5035)
                    (0.11, 0.6033)
                    (0.12, 0.6878)
            	};
            	\addplot[solid, color=orange, mark=triangle*] plot coordinates {    
                	(0.05, 0.0321)
                    (0.06, 0.0805)
                    (0.07, 0.1580)
                    (0.08, 0.2627)
                    (0.09, 0.3823)
                    (0.10, 0.5059)
                    (0.11, 0.6143)
                    (0.12, 0.7042)
            	};
            	\addplot[solid, color=olive, mark=star] plot coordinates {
                	(0.05, 0.0158)
                    (0.06, 0.0492)
                    (0.07, 0.1186)
                    (0.08, 0.2298)
                    (0.09, 0.3629)
                    (0.10, 0.5036)
                    (0.11, 0.6318)
                    (0.12, 0.7332)
            	};
            	\legend{$d=6$,$d=8$,$d=9$,$d=12$}
            \end{semilogyaxis}    
      	\end{tikzpicture}
  	}
    \subcaption{\label{fig:HInvNNDec}}
\end{subfigure}
\begin{subfigure}[b]{0.48\textwidth}
    \centering
    \resizebox{\linewidth}{!}{
    	\begin{tikzpicture}
            \begin{semilogyaxis}[
                xlabel={$p_{err}$}, 
                ylabel={Logical error}, 
                legend pos=south east,
                every axis x label/.style={
                    at={(ticklabel* cs:1.025)},
                    anchor=north west
                },
                ymin=1e-2, ymax=1e0
            ]
            	\addplot[solid, color=blue, mark=*] plot coordinates {    
                	(0.05, 0.0874)
                    (0.06, 0.1483)
                    (0.07, 0.2254)
                    (0.08, 0.3102)
                    (0.09, 0.3966)
                    (0.10, 0.4841)
                    (0.11, 0.5636)
                    (0.12, 0.6355)
            	};
            	\addplot[solid, color=red, mark=square*] plot coordinates {    
                	(0.05, 0.0394)
                    (0.06, 0.0883)
                    (0.07, 0.1615)
                    (0.08, 0.2556)
                    (0.09, 0.3623)
                    (0.10, 0.4717)
                    (0.11, 0.5757)
                    (0.12, 0.6641)
            	};
            	\addplot[solid, color=orange, mark=triangle*] plot coordinates {    
                	(0.05, 0.0237)
                    (0.06, 0.0615)
                    (0.07, 0.1289)
                    (0.08, 0.2235)
                    (0.09, 0.3383)
                    (0.10, 0.4602)
                    (0.11, 0.5750)
                    (0.12, 0.6730)
            	};
            	\addplot[solid, color=olive, mark=star] plot coordinates {
                	(0.05, 0.0122)
                    (0.06, 0.0388)
                    (0.07, 0.0971)
                    (0.08, 0.1967)
                    (0.09, 0.3232)
                    (0.10, 0.4656)
                    (0.11, 0.5969)
                    (0.12, 0.7085)
            	};
            	\legend{$d=6$,$d=8$,$d=9$,$d=12$}
            \end{semilogyaxis}    
      	\end{tikzpicture}
  	}
    \subcaption{\label{fig:HInvNNDecTwo}}
\end{subfigure}
\caption{\label{fig:performanceFigs} The performance of neural decoder in first approach, achieving a threshold of $10 \%$ is shown in (a). The performance of neural decoder in second approach, achieving a near optimal threshold is shown in (b). Note the reduction in logical error for decoder in second approach (b) when compared to that of first approach (a).}
\end{figure*}

The performance of our neural decoder in first approach (Fig.~\ref{fig:decoderFlowDigOne}) trained according to the training procedure mentioned in Section~\ref{subsec:OurWork:TrainingProcedure} is shown in the Fig.~\ref{fig:HInvNNDec}. The fully trained NN model is independent of the $p_{err}$ and the it outperforms the previous state-of-the art methods which are not based on neural networks by~\cite{sarvepalli12,delfosse14,bombin2012universal}. We report that our neural decoder achieves a threshold of $10\%$ and is comparable to the result mentioned in \cite{maskara2018advantages}.

\begin{figure*}
\includegraphics[scale=0.55]{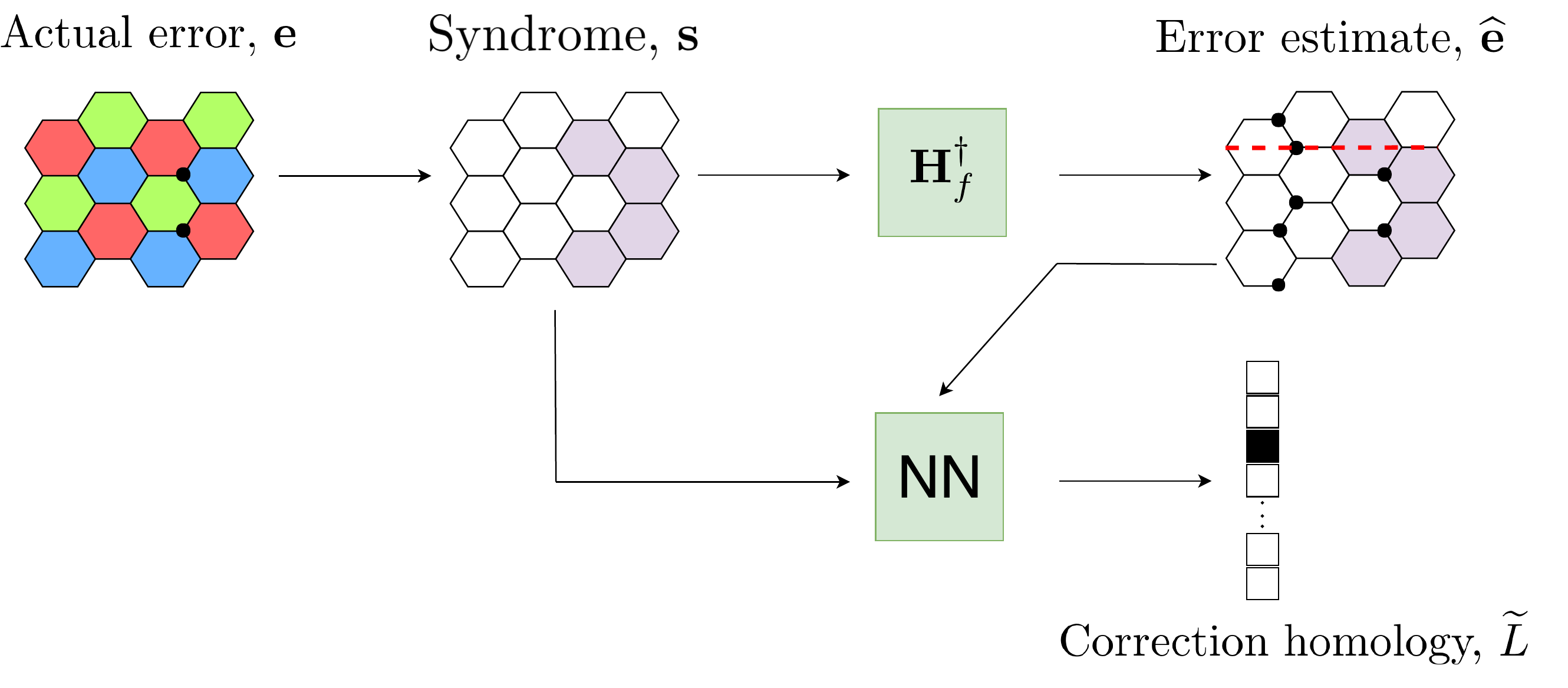}
\caption{\label{fig:decoderFlowDigTwo} Flow diagram of our two-step decoder. The black dots represent error on the qubits and the marked regions represent the syndrome caused. In the first-step we get an estimate of the error $\mathbf{\widehat{e}}$ and in the second-step, we predict the correction homology $\widetilde{L}$ using our trained NN with the information of both $\widehat{\mathbf{e}}$ and $\mathbf{s}$. Our final error correction is same as $\widetilde{L} \widehat{E}$.}
\end{figure*}

In our second approach, we have given additional information of $\mathbf{\widehat{e}}$ along with the syndrome vector $\mathbf{s}$ (by concatenating them both) to our NN (Fig.~\ref{fig:decoderFlowDigTwo}) and saw a dramatic improvement in the threshold for small lengths, as well as a reduction in logical errors for each error rate as shown in the Fig.~\ref{fig:HInvNNDecTwo}. The training is exactly similar to the previous case.
This shows that the NN is able to understand and learn the behaviour of the $\mathbf{H}$-inverse decoder much better with the additional knowledge of the initial estimate $\mathbf{\widehat{e}}$ and hence is able to perform better correction. This implies that the data driven methods and in particular neural networks' performance can be improved by providing all the information available to us relevant to the problem to be solved. This modification can be incorporated into other works of building two-step decoders using neural networks and improve the overall performance. 

The hyper-parameters (as described in the Section~\ref{subsec:Background:ML_DL}) of our networks are listed in the Tables~\ref{tab:hyperParameters},~\ref{tab:hyperParametersTwo} for first and second approaches respectively. The variation of some of them with the distance $d$ are shown in the Fig.~\ref{fig:HPvsD} for both the approaches. The distance of the code is denoted by $d$ and the number of hidden layers in our network is denoted by $h_{d}$. The batch size used for each length is denoted by $b_{d}$. The number of nodes in each hidden layer are characterized by the hidden dimension factor $f_{d}$ which is equal to $f_{d}$ multiplied by the dimension of the input syndrome vector $\mathbf{s}$. The parameter $t_{d, p_{err}}$ is the number of samples required for training for each $p_{err}$ and $T_{d}$ determines the total number of samples the final trained NN has seen entirely. The parameter $\alpha$ is the learning rate used for optimization. We used \textit{PyTorch} \footnote{\url{https://pytorch.org/}}, an open-source deep learning framework for training our neural networks.

\begin{table}
\caption{\label{tab:hyperParametersTwo}The values of the hyper-parameters used in the neural decoder in our second approach.}
\begin{ruledtabular}
\begin{tabular}{c|cccccc}
\diagbox{$d\footnote{Distance of the code}$}{parameters} & $h_{d}\footnote{Number of hidden layers}$ & $f_{d}\footnote{Hidden dimension factor}$ & $b_{d}\footnote{Batch size}$ & $\alpha\footnote{Learning rate}$ & $t_{d, p_{err}}\footnote{Number of training samples per each $p_{err}$}$ & $T_{d}\footnote{Total number of training samples for all $p_{err}$ combined}$\\
\hline
$6$ & $1$ & $1$ & $500$ & $0.001$ & $2 \times 10^{7}$ & $1.4 \times 10^{8}$\\
$8$ & $2$ & $3$ & $750$ & $0.001$ & $4 \times 10^{7}$ & $2.8 \times 10^{8}$\\
$9$ & $3$ & $4$ & $750$ & $0.001$ & $4 \times 10^{7}$ & $2.8 \times 10^{8}$\\
$12$ & $6$ & $10$ & $2500$ & $0.001$ & $10 \times 10^{7}$ & $7 \times 10^{8}$\\
\end{tabular}
\end{ruledtabular}
\end{table}

\section{\label{sec:Insights} Remarks and Insights}
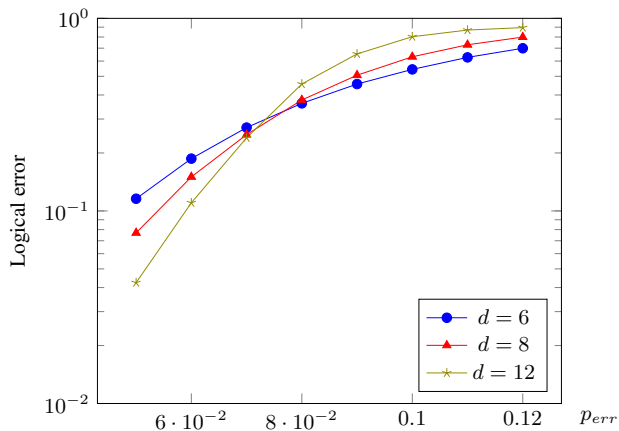
\begin{figure}
\centering
\begin{tikzpicture}[scale=0.9]
\begin{semilogyaxis}[
    xlabel={$p_{err}$}, 
    ylabel={Logical error}, 
    legend pos=south east,
    every axis x label/.style={
        at={(ticklabel* cs:1.025)},
        anchor=north west
    },
    ymin=1e-2, ymax=1e0
]
	\addplot[solid, color=blue, mark=*] plot coordinates {
    	(0.05, 0.1157)
        (0.06, 0.1869)
        (0.07, 0.2709)
        (0.08, 0.3619)
        (0.09, 0.4559)
        (0.10, 0.5432)
        (0.11, 0.6270)
        (0.12, 0.6993)
	};
	\addplot[solid, color=red, mark=triangle*] plot coordinates {
    	(0.05, 0.0769)
        (0.06, 0.1502)
        (0.07, 0.2489)
        (0.08, 0.3768)
        (0.09, 0.5067)
        (0.10, 0.6315)
        (0.11, 0.7290)
        (0.12, 0.7993)
	};
	\addplot[solid, color=olive, mark=star] plot coordinates {
    	(0.05, 0.0424)
        (0.06, 0.1102)
        (0.07, 0.2397)
        (0.08, 0.4561)
        (0.09, 0.6535)
        (0.10, 0.8029)
        (0.11, 0.8694)
        (0.12, 0.8952)
	};
	\legend{$d=6$,$d=8$,$d=12$}
\end{semilogyaxis}
\end{tikzpicture}
\caption{\label{fig:nonProgressiveNNDec} Performance of our neural decoder without the progressive training procedure. The threshold achieved is just about $7.2 \%$.}
\end{figure}

We clearly demonstrate the power of data-driven methods and in particular neural networks, through which we were able to improve the performance of a very bad decoder which does not even have a threshold. 
When compared to the previous state-of-the-art on neural decoders for color codes, our decoder requires significantly less training data for higher lengths like $d=9, 12$. 
In addition to the gains in training cost, our decoder has less complexity with respect to the number of layers and number of nodes in each layer when compared to the previous work and still achieved a comparable threshold.
In Section~\ref{subsec:OurWork:TrainingProcedure}, we mentioned the importance of the progressive training. We ran our simulations by training a new NN with Xavier normal and Gaussian normal initializations for every $p_{err}$, without employing the progressive training. The performance of that decoder with similar hyper-parameters as mentioned in the Table~\ref{tab:hyperParameters} is shown in the Fig.~\ref{fig:nonProgressiveNNDec}. This shows that without the progressive training, the threshold of the decoder drops to about $7.2 \%$. This is because as the $p_{err}$ increases, it would be very likely that our optimizer converges to a bad local minima. This progressive training is similar to the common practice of \textit{curriculum-learning} in neural networks so that the optimizer converges to a better local minima in the hyperspace of the network weights as proposed in~\cite{Bengio:2009:CL:1553374.1553380}. 
We also report that this progressive training should be carried on till the $p_{err}$ equals the theoretical threshold and we have observed constant decrement in logical errors at all error rates. Training the model with a $p_{err}$ above the threshold is not desirable as we have seen increments in the logical errors. 
This concept of $\mathbf{H}$-inverse as a base decoder improved with a neural decoder can be effectively extended to other noise models and also to codes in higher dimension including other stabilizer codes. 

Any decoder which does error correction essentially solves the equation $\mathbf{H} \mathbf{x}^{\top} = \mathbf{s}$. 
Since there are many solutions, it implies there exist many pseudo-inverses to $\mathbf{H}$. 
To implement a good decoder, choosing the correct inverse for a given syndrome is an important task. 
Different inverses must be chosen for different syndrome patterns in order to have a threshold. 
The choice of decoder in the step-one can be anything as long as it clears the syndrome and good decoders which have a threshold can also be chosen. 
In such cases, these good decoders take care of selecting the inverse depending on the syndrome. 
This makes these step-one decoders not entirely simple and there is a lot more for the NN to learn to improve the initial estimate.
This is because the inverse selected will be different for different syndromes. 
In our approach, we fix the inverse $\mathbf{H}_{f}^{\dagger}$ though it does not have a threshold and make the step-one decoder very simple. 
Our NN only has to understand on inverse which is $\mathbf{H}_{f}^{\dagger}$ to improve the initial estimate. 
Intuitively, this means that the learning should be easier for our NN which can be verified empirically through the superior performance with comparatively lesser training cost and complexity when compared to \cite{maskara2018advantages}. 
Our approach is applicable for any decoding problem where the equation, $\mathbf{H} \mathbf{x}^{\top} = \mathbf{s}$ needs to be solved.

\section{\label{sec:Conclusion} Conclusion}
We have demonstrated that data-driven methods like NN can perform superior decoding when compared to the traditional approaches.
We propose a neural decoder with simplified non-neural part achieving a threshold of $10 \%$ for 2D color codes.
We suggest an alternative approach to combine non-neural and neural decoders reducing the logical error which can be incorporated into other NN based decoders.
The drawbacks of NN based decoders are figuring out the right set of hyper-parameters for each length and practical issues of convergence of the loss when the number of trainable parameters increase.
Our approach can be extended to other realistic noise models and codes in higher dimensions or other stabilizer codes.

\section{\label{sec:Acknowledgements} Acknowledgements}
The authors would like to thank Arun B. Aloshious for valuable discussions. During the preparation of this manuscript, five related preprints were made available \cite{ni2018neural,sweke2018reinforcement,liu2018neural,andreasson2018quantum,nautrup2018optimizing}, however their scope and emphasis are different from our work. This work was completed when CC was associated with Indian Institute of Technology Madras as a part of his Dual Degree thesis.

\bibliography{aipsamp}

\end{document}